\documentclass[a4paper,aps,prd,10pt,preprintnumbers,showpacs,
twocolumn,superscriptaddress,nofootinbib,amsmath,amssymb]{revtex4-1}
\usepackage{setspace}
\usepackage{graphicx,multirow,cmap,ulem}
\usepackage[utf8]{inputenc}
\usepackage[T1]{fontenc}
\usepackage{hyperref}
\usepackage{soul}

\def\re#1{Re(#1)}
\def\im#1{Im(#1)}
\def\Order#1{{\cal O}\left(#1\right)}

\begin{document}

\title{Long-lived quasinormal modes, shadows and particle motion in four-dimensional quasi-topological gravity}
\author{Bekir Can Lütfüoğlu}
\email{bekir.lutfuoglu@uhk.cz}
\affiliation{Department of Physics, Faculty of Science, University of Hradec Králové, Rokitanského 62/26, 500 03 Hradec Králové, Czech Republic}
\begin{abstract}
We investigate massive scalar perturbations and several characteristics of particle motion in the spacetime of regular black holes arising in four-dimensional quasi-topological gravity. Quasinormal modes are computed using high-order WKB approximations with Padé resummation and verified through time-domain integration. For moderate values of the scalar-field mass, the time-domain profiles confirm the WKB results with excellent accuracy. As the mass increases, the damping rate decreases substantially, indicating the approach to the quasi-resonant regime of long-lived modes. For sufficiently large masses, the late-time signal becomes dominated by oscillatory power-law tails, which mask the quasi-resonant mode in the time-domain profile. In addition, we analyze photon motion and circular geodesics, including the photon-sphere radius, shadow size, Lyapunov exponent, and ISCO characteristics. These quantities exhibit only moderate deviations from their Schwarzschild values, unlike the Hawking temperature of the black hole. 
\end{abstract}
\maketitle

\section{Introduction}

The exterior of black-holes predicted by classical general relativity is described by the Schwarzschild or Kerr spacetimes, and these geometries do not contradict recent observations in gravitational and electromagnetic spectra \cite{LIGOScientific:2016aoc,LIGOScientific:2017vwq,LIGOScientific:2020zkf,EventHorizonTelescope:2019dse}. However, these solutions of classical Einstein equations contain spacetime singularities where curvature invariants diverge and geodesic evolution becomes incomplete. The presence of such regions signals the breakdown of the classical description of gravity and indicates that the geometry must be modified in the strong–field regime. One possible resolution of this difficulty is provided by regular black holes, whose central region is replaced by a geometry with finite curvature invariants.

Most regular black-hole solutions were obtained via a kind of ad hoc procedure. This was achieved either by coupling Einstein gravity to some matter sources or by introducing effective models motivated by quantum gravity considerations \cite{Bardeen:1968, Dymnikova:1992ux, AyonBeato:1998ub, Bonanno:2000ep, Bronnikov:2000vy, Bronnikov:2005gm,  Hayward:2005gi, Ansoldi:2008jw, Dymnikova:2015yma, Bronnikov:2024izh, Spina:2025wxb,  Bolokhov:2025zva, Bolokhov:2025fto, Konoplya:2025ect}. In such cases, the required matter content often does not necessarily arise from a fundamental gravitational theory. An appealing alternative is that regular black holes arise as vacuum solutions of modified gravity theories that include higher–curvature corrections.

One of the recent findings in this area is the construction of regular black holes in four-dimensional non-polynomial quasi-topological gravity \cite{Borissova:2026wmn, Bueno:2025zaj}. Within this framework, black-hole solutions can be obtained analytically. The resulting geometries are asymptotically flat and free of curvature singularities, reproducing well-known regular metrics without introducing additional matter sources. A regularization parameter usually controls the deviation from the Schwarzschild spacetime and is typically localized in the near-horizon region.

Understanding how such geometrically regular black holes manifest themselves through observable phenomena requires analyzing both wave dynamics and particle motion in these spacetimes. One of the most important probes of black-hole geometry is provided by {\it quasinormal modes} (QNMs), which describe the characteristic oscillations of perturbed black-holes \cite{Konoplya:2011qq, Bolokhov:2025rng, Kokkotas:1999bd}. These modes dominate the ringdown stage of gravitational-wave signals produced in black-hole mergers and therefore play a key role in extracting information about the underlying black-hole spacetime geometry \cite{LIGOScientific:2016aoc, LIGOScientific:2017vwq, LIGOScientific:2020zkf}. At the same time, properties of geodesic motion, such as the angular velocity at the innermost stable circular orbit (ISCO), the Lyapunov exponent governing the instability of photon orbits, the binding energy of particles, and the radius of the black-hole shadow, provide complementary observational diagnostics that directly probe the structure of the gravitational field in the strong-curvature regime.

Quasinormal modes of {\it massive} fields have attracted considerable attention due to a number of interesting spectral features that arise when the field mass is varied (see, for example, \cite{Konoplya:2004wg, Konoplya:2018qov, Zhidenko:2006rs, Konoplya:2017tvu, Konoplya:2005hr, Ohashi:2004wr, Zhang:2018jgj, Aragon:2020teq, Ponglertsakul:2020ufm, Gonzalez:2022upu, Ponglertsakul:2020ufm, Burikham:2017gdm, Bolokhov:2023ruj} and references therein). Massive perturbations appear naturally in a variety of theoretical contexts. For instance, effective mass terms may arise in higher-dimensional scenarios due to the influence of the bulk on brane-localized fields \cite{Seahra:2004fg}. In addition, massive gravitons or effective massive modes can contribute to the spectrum of very long gravitational waves, which are currently being investigated by Pulsar Timing Array experiments \cite{Konoplya:2023fmh, NANOGrav:2023hvm}. 

A particularly intriguing feature of massive perturbations is the possibility of arbitrarily long-lived modes when the field mass approaches certain critical values \cite{Ohashi:2004wr, Konoplya:2004wg}. This phenomenon has been observed for different spins of the perturbing field \cite{Konoplya:2005hr, Fernandes:2021qvr, Konoplya:2017tvu, Percival:2020skc}, various black-hole backgrounds \cite{Zhidenko:2006rs, Zinhailo:2018ska, Churilova:2020bql, Bolokhov:2023bwm}. The presence of the mass term also qualitatively modifies the late-time decay of perturbations: instead of the usual power-law tails characteristic of massless fields, the signal exhibits oscillatory late-time behavior \cite{Jing:2004zb, Koyama:2001qw, Moderski:2001tk, Rogatko:2007zz, Koyama:2001ee, Koyama:2000hj, Gibbons:2008gg, Gibbons:2008rs}. In certain situations, an effective mass term may also appear dynamically, for example, when a massless field propagates in the vicinity of a black hole immersed in a magnetic field \cite{Konoplya:2007yy, Konoplya:2008hj, Wu:2015fwa}. At the same time, the existence of arbitrarily long-lived modes is not universal, and there are examples where such modes do not occur \cite{Zinhailo:2024jzt, Konoplya:2005hr}. However, the behavior of massive perturbations and their associated observational characteristics have not yet been studied for regular black holes arising in four-dimensional quasi-topological gravity.

In this work, we investigate the quasinormal spectrum of a massive scalar field and several characteristics of particle motion in the spacetime of regular black holes arising in four-dimensional quasi-topological gravity. While the massless perturbations and QNMs have been recently studied in \cite{Konoplya:2026gim}, no such analysis exists for massive perturbations. In particular, here we analyze the dependence of the quasinormal spectrum on the mass of the field and study how the regularization of the geometry influences geodesic observables such as the Lyapunov exponent of unstable photon orbits, the angular velocity at the ISCO, the binding energy of particles, and the black-hole shadow. These quantities provide complementary probes of the spacetime geometry and allow one to assess how the near-horizon modifications characteristic of regular black holes affect both wave dynamics and particle motion.

The paper is organized as follows. In Sec.~\ref{sec:SecII}, we briefly review the regular black-hole geometries arising in four-dimensional quasi-topological gravity and discuss the metric functions used in the present analysis. In Sec.~\ref{sec:SecIII}, we derive the wave equation for a massive scalar field and formulate the QNM boundary conditions. Section~\ref{sec:SecIV} describes the methods employed to compute the spectrum, namely the high-order WKB approximation with Padé resummation and the time-domain integration scheme. The obtained QNMs and their dependence on the scalar-field mass are presented and analyzed in Sec.~\ref{sec:SecV}. Then, in Sec.~\ref{sec:SecVI}, we investigate several characteristics of particle motion, including photon-sphere properties, black-hole shadows, Lyapunov exponents, and the parameters of the ISCO. Finally, Sec.~\ref{sec:SecVII} summarizes the main results and conclusions of the paper.

\section{Regular black-hole geometries in quasi-topological gravity}\label{sec:SecII}

Non-polynomial quasi-topological gravity (NP-QTG) represents a class of modified gravitational theories in which higher-curvature invariants are combined in such a way that the resulting dynamics remains well behaved in symmetric backgrounds \cite{Bueno:2024dgm, Bueno:2024eig, Bueno:2025tli, Frolov:2024hhe}. The action of the theory can be written schematically as \cite{Bueno:2025zaj, Borissova:2026wmn}
\begin{equation}
S = \frac{1}{16\pi G}\int d^4x\,\sqrt{-g} \left[ R + \mathcal{F}(\mathcal{R}) \right],
\end{equation}
where $\mathcal{R}$ denotes a particular curvature scalar constructed from contractions of the Riemann tensor and its traces, while $\mathcal{F}$ is a non-polynomial function of this invariant. The function $\mathcal{F}$ is chosen so that the theory remains ghost-free around maximally symmetric backgrounds and the equations governing static and spherically symmetric configurations simplify considerably. In particular, the trace of the field equations reduces to a total derivative in the spherically symmetric sector, which makes it possible to construct exact black-hole solutions in closed analytic form.

We consider static and spherically symmetric geometries described by the line element
\begin{equation}
ds^{2}=-f(r)dt^{2}+\frac{dr^{2}}{f(r)}+r^{2}d\Omega^{2},
\end{equation}
where the entire structure of the spacetime is determined by the metric function $f(r)$. Within the NP-QTG framework the field equations for this ansatz reduce to a single algebraic relation for $f(r)$, allowing families of regular black-hole solutions to be obtained explicitly.

A broad class of such geometries can be written in the form \cite{Tsuda:2026xjc, Konoplya:2026gim}
\begin{equation}
f(r)=1- \frac{2Mr^{\mu-1}} {\left(r^{\nu}+\alpha^{\nu/3}(2M)^{\nu/3}\right)^{\mu/\nu}},
\end{equation}
where $M$ is the integration constant associated with the ADM mass of the black hole and $\alpha$ is a parameter with dimensions of length squared controlling the deviation from the Schwarzschild solution. For convenience, we denote $\alpha=l^{2}$ and measure all dimensional quantities in units of the black-hole mass $M$. Different choices of the integers $\mu$ and $\nu$ correspond to different regular metrics belonging to this family.

In the present work, we focus on two representative examples of regular black-hole geometries obtained within quasi-topological gravity. These metrics illustrate typical features of the theory while remaining sufficiently simple for detailed analysis.

\medskip
\noindent
{\bf Model I.}
For particular values of the parameters, $\mu=3$ and $\nu=6$, in the general expression above one obtains the metric function
\begin{equation}
f(r)= 1 - \frac{2Mr^{2}}{\sqrt{4l^{4}M^{2}+r^{6}}},
\end{equation}
which describes a regular black hole emerging from the quasi-topological gravity framework \cite{Borissova:2026wmn}. The parameter $l$ controls the strength of the near-core modification of the geometry: when $l$ increases, the deviation from the Schwarzschild metric becomes more pronounced in the strong-field region.

\medskip
\noindent
{\bf Model II.}
As a second example, we consider another regular solution discussed in \cite{Borissova:2026wmn}, whose metric function takes the form
\begin{equation}
f(r)=1-\frac{r^2}{l^2}\left(1-e^{-\frac{l^2 M}{r^3}}\right).
\end{equation}
Although this solution does not belong to the algebraic family described above, it arises within the same quasi-topological gravity framework and shares similar qualitative properties.

While there exists a large freedom in the choice of generating functions leading to a wide variety of black hole solutions—many of which can be regular—a few particularly representative metrics, such as those of Hayward \cite{Hayward:2005gi}, Dymnikova \cite{Dymnikova:1992ux,Dymnikova:2015yma},  Bardeen \cite{Bardeen:1968} and related models, have been extensively investigated in the literature \cite{Fernando:2012yw,Saleh:2018hba,Konoplya:2023ppx,Konoplya:2023aph,Lutfuoglu:2026fpx,Lutfuoglu:2025pzi,Konoplya:2022hll,Lutfuoglu:2025ohb}. In the present work, we consider a new regular black hole metric possessing a set of desirable properties, which will be specified below.

We demonstrate that, when analyzed alongside previously studied cases, this solution exhibits a number of common features at the level of observable quantities. In particular, higher-curvature corrections tend to produce qualitatively similar effects across different models. This universality enhances the significance of our results: although specific black hole constructions may differ, one can expect broadly similar spectral and orbital characteristics for black holes modified by higher-curvature contributions.

\medskip

Both geometries are asymptotically flat and reduce to the Schwarzschild solution in the limit $l\to0$. At the same time, the spacetime remains regular at the center: the metric function is finite at $r=0$, and the curvature invariants do not diverge. The corrections to the Schwarzschild geometry are primarily concentrated in the vicinity of the event horizon, while at large distances the metrics approach the standard vacuum solution of general relativity. Consequently, one expects only moderate modifications of the effective potentials governing wave propagation and geodesic motion, making these spacetimes suitable laboratories for investigating how regularization of the central region affects observable properties such as QNMs and particle dynamics.

\section{Massive scalar perturbations and quasinormal boundary conditions} \label{sec:SecIII}

We consider a test scalar field of mass $\mu$ propagating in the background of a static, spherically symmetric, and asymptotically flat black hole described by
\begin{equation}
ds^{2}=-f(r)dt^{2}+\frac{dr^{2}}{f(r)}+r^{2}\left(d\theta^{2}+\sin^{2}\theta\,d\phi^{2}\right).
\end{equation}
The field obeys the Klein--Gordon equation
\begin{equation}\label{KGmassive}
\left(\Box-\mu^{2}\right)\Phi=0,
\end{equation}
where $\mu$ is the mass of the scalar field.

Using the standard separation of variables (see, for instance \cite{Bagrov:1990hv, Konoplya:2018arm} and references therein),
\begin{equation}
\Phi(t,r,\theta,\phi)=e^{-i\omega t}Y_{\ell m}(\theta,\phi)\frac{\Psi(r)}{r},
\end{equation}
where $Y_{\ell m}(\theta,\phi)$ are the spherical harmonics, one reduces Eq.~(\ref{KGmassive}) to a radial wave equation. Introducing the tortoise coordinate $r_*$ by
\begin{equation}\label{tortoise_massive}
\frac{dr_*}{dr}=\frac{1}{f(r)},
\end{equation}
the radial part takes the Schr\"odinger-like form
\begin{equation}\label{wave_massive}
\frac{d^{2}\Psi}{dr_*^{2}}+\left(\omega^{2}-V(r)\right)\Psi=0.
\end{equation}

For a massive scalar field, the effective potential is given by
\begin{equation}\label{Vmassive}
V(r)=f(r)\left(\frac{\ell(\ell+1)}{r^{2}}+\frac{f'(r)}{r}+\mu^{2}\right).
\end{equation}
Thus, compared with the massless case, the scalar-field mass contributes an additional term $f(r)\mu^{2}$, which raises the effective potential and qualitatively changes its asymptotic behavior.

\begin{figure*}
\resizebox{\linewidth}{!}{\includegraphics{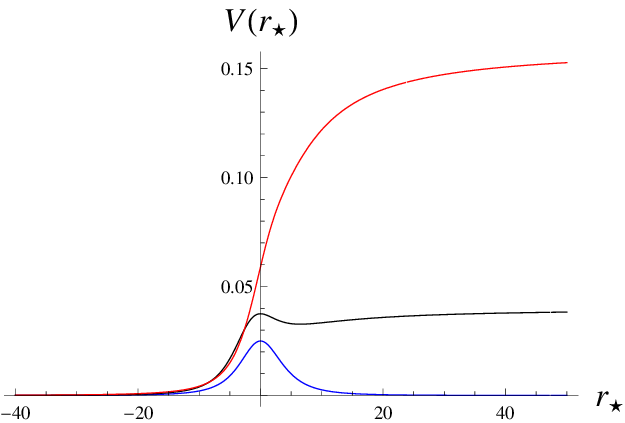}\includegraphics{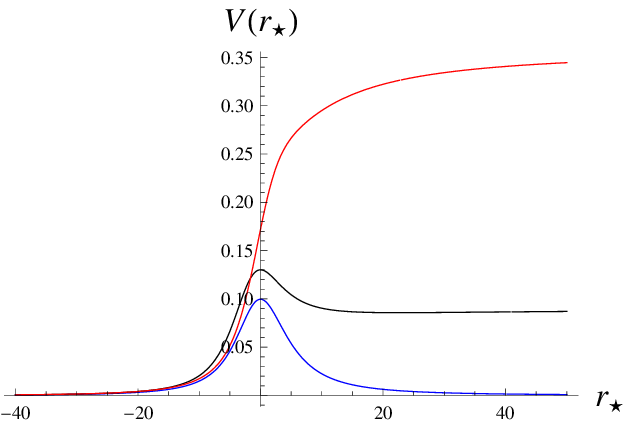}}
\caption{Black hole model I. Effective potential as a function of the tortoise coordinate $r^{*}$ for $\ell=0$ (left panel: $\mu=0$ (blue), $\mu=0.2$ (black), and $\mu=0.4$ (red)) and $\ell=1$ (right panel: $\mu=0$ (blue), $\mu=0.3$ (black), and $\mu=0.6$ (red)). The parameters are $4l^{4}=0.94$ and $M=1$.}\label{fig:potL01}
\end{figure*}

\begin{figure*}
\resizebox{\linewidth}{!}{\includegraphics{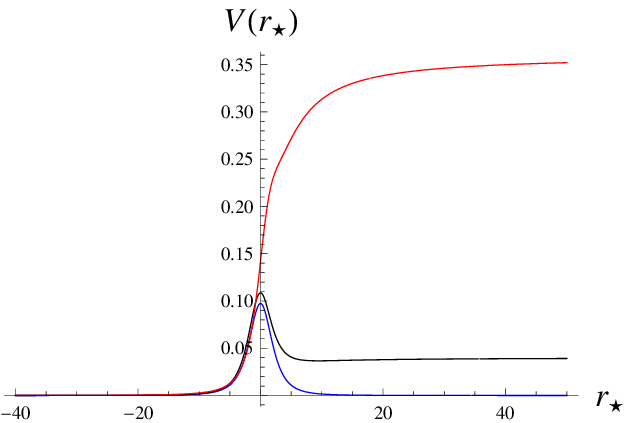}\includegraphics{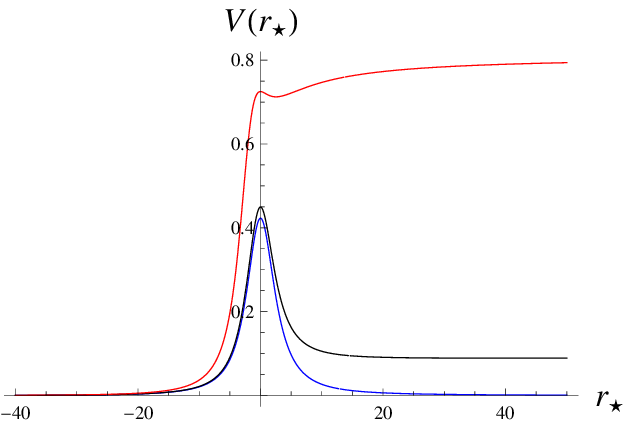}}
\caption{Black hole model II. Effective potential as a function of the tortoise coordinate $r^{*}$ for $\ell=0$ (left panel: $\mu=0$ (blue), $\mu=0.2$ (black), and $\mu=0.6$ (red)) and $\ell=1$ (right panel: $\mu=0$ (blue), $\mu=0.3$ (black), and $\mu=0.9$ (red)). The parameters are $l=0.51$ and $M=1$.}\label{fig:potL01Model2}
\end{figure*}

The QNM boundary conditions are
\begin{equation}\label{bc_massive}
\Psi(r_*)\sim
\begin{cases}
e^{-i\omega r_*}, & r_*\to -\infty,\\[4pt]
e^{+i\chi r_*}, & r_*\to +\infty,
\end{cases}
\end{equation}
where
\begin{equation}\label{chi_massive}
\chi=\sqrt{\omega^{2}-\mu^{2}}.
\end{equation}
The first condition corresponds to a purely ingoing wave at the event horizon, while the second one imposes the absence of incoming radiation from spatial infinity. The branch of the square root is chosen so that the asymptotic solution represents an outgoing wave, i.e., $\re{\chi}$ and $\re{\omega}$ have the same sign. When $\omega^{2}$ approaches $\mu^{2}$, the quantity $\chi$ becomes small, and the damping rate may decrease substantially, leading to the well-known long-lived modes and quasiresonances of massive fields.

It is worth emphasizing that the scalar mass modifies not only the asymptotic value of the potential but also its overall shape, as illustrated in Figs.~\ref{fig:potL01} and ~\ref{fig:potL01Model2}. As $\mu$ increases, the height and width of the barrier are altered, and for sufficiently large $\mu$ the potential may cease to possess a well-defined maximum for some values of the multipole number $\ell$. In that regime, the standard WKB treatment is no longer applicable, and the spectrum must be analyzed with more general methods. For the range of parameters where the potential retains a single-barrier form, however, Eq.~(\ref{wave_massive}) together with the boundary conditions~(\ref{bc_massive}) provide the basis for the computation of quasinormal frequencies.

\section{Methods of calculation} \label{sec:SecIV}

In order to determine the quasinormal spectrum of the massive scalar field, we employ two complementary approaches: the WKB approximation and direct time-domain integration. The former provides accurate values of quasinormal frequencies when the effective potential has a single barrier shape, while the latter allows one to verify the results by studying the dynamical evolution of perturbations.

\subsection{WKB approximation}

The WKB method is based on matching the asymptotic solutions of the wave equation
\begin{equation}
\frac{d^{2}\Psi}{dr_*^{2}}+\left(\omega^{2}-V(r)\right)\Psi=0,
\end{equation}
with the Taylor expansion of the effective potential near its maximum. For black-hole perturbations whose effective potential forms a single barrier, this approach yields an approximate quantization condition for the quasinormal frequencies.

At leading orders the WKB formula can be written as \cite{Schutz:1985km, Iyer:1986np, Konoplya:2003ii, Matyjasek:2017psv, Konoplya:2019hlu}
\begin{equation}\label{WKBformula}
\omega^{2}=V_{0}-i\left(n+\frac12\right)\sqrt{-2V_{0}''}
+\sum_{k=2}^{\infty}\Lambda_k ,
\end{equation}
where $V_0$ is the value of the effective potential at its maximum, $V_0''$ is the second derivative of the potential with respect to the tortoise coordinate evaluated at the same point, and $n=0,1,2,\ldots$ denotes the overtone number. The terms $\Lambda_k$ represent higher-order WKB corrections which depend on higher derivatives of the potential at the maximum.

In practice, the WKB approximation can be pushed to high orders and combined with Padé resummation \cite{Matyjasek:2017psv, Matyjasek:2019eeu}, which significantly improves the accuracy of the method. The WKB approach is particularly efficient for low overtones and moderate or large multipole numbers, where the effective potential is well approximated by a single smooth barrier.

In this work, we make use of high-order WKB approximations following the formulations of Refs.~\cite{Matyjasek:2019eeu, Matyjasek:2017psv}. For backgrounds that admit analytic treatment, we employ the 14th- and 16th-order expressions. A useful practical criterion of reliability is the stability of the result with respect to the WKB order: when several neighboring orders produce nearly the same quasinormal frequency, the expected uncertainty is typically comparable to the difference between these consecutive approximations. At the same time, it should be kept in mind that the WKB expansion represents an asymptotic series, so that increasing the order does not necessarily lead to a strictly monotonic improvement of accuracy. The WKB technique has been widely elaborated and employed in the analysis of black hole perturbations (see, for instance, \cite{Konoplya:2006ar, Malik:2024tuf, Hamil:2024njs, Skvortsova:2024atk, Han:2026fpn, Konoplya:2025hgp, Malik:2025erb, Kodama:2009bf, Lutfuoglu:2025pzi, Arbelaez:2026eaz, Konoplya:2023moy, Bolokhov:2024ixe, Lutfuoglu:2026xlo, Konoplya:2009hv, Kanti:2006ua, Bolokhov:2023dxq, Lutfuoglu:2025hjy, Skvortsova:2023zmj, Bolokhov:2024bke, Lutfuoglu:2025eik, Stuchlik:2025ezz, Bolokhov:2025egl}), and therefore we refrain from repeating a detailed description of the method here.

\subsection{Time-domain integration}

To verify the quasinormal frequencies obtained with the WKB method, we also study the dynamical evolution of perturbations in the time domain. For this purpose, we integrate the wave equation using the characteristic discretization scheme introduced by Gundlach, Price and Pullin \cite{Gundlach:1993tp}.

Introducing the null coordinates
\begin{equation}
u=t-r_*, \qquad v=t+r_*,
\end{equation}
the wave equation takes the form
\begin{equation}
4\frac{\partial^{2}\Psi}{\partial u \partial v}+V(r)\Psi=0.
\end{equation}
Discretizing the $(u,v)$ plane on a numerical grid, the field values are obtained using the relation
\begin{eqnarray}\label{GPP}
\Psi(N)&=&\Psi(W)+\Psi(E)-\Psi(S) \nonumber\\
&&-\frac{\Delta^{2}}{8}V(S)\left[\Psi(W)+\Psi(E)\right]
+\mathcal{O}(\Delta^{4}),
\end{eqnarray}
where $N=(u+\Delta,v+\Delta)$, $W=(u+\Delta,v)$, $E=(u,v+\Delta)$ and $S=(u,v)$ denote the grid points of the characteristic integration scheme.

\begin{figure*}
\resizebox{\linewidth}{!}{\includegraphics{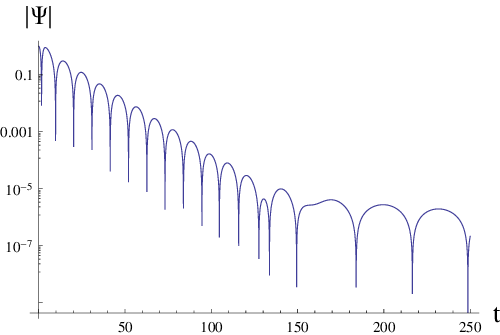}\includegraphics{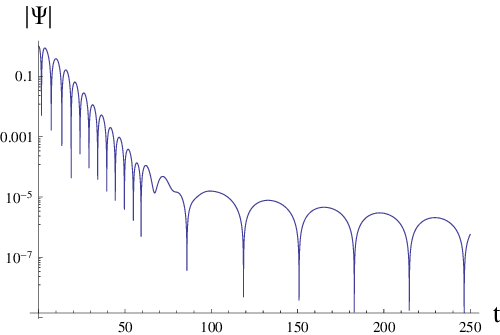}}
\caption{Right Panel: Black hole model I. Time-domain profile for $\ell=1$ perturbations. The parameters are $l=0.51$ and $M=1$, $\mu=0.1$. The time-domain integration dominant mode is $\omega =0.29641 - 0.087796  i$, while the WKB data is $\omega = 0.296401 - 0.0877942 i$, confirming great accuracy of the WKB method.  Left Panel: Black hole model II. Time-domain profile for $\ell=1$ perturbations. The parameters are $l=0.51$ and $M=1$, $\mu=0.1$. The time-domain integration dominant mode is $\omega =0.612748 - 0.166874 i$, while the WKB data is $\omega = 0.612667 -0.166952 i$.}\label{fig:TD2}
\end{figure*}

\begin{figure}
\resizebox{\linewidth}{!}{\includegraphics{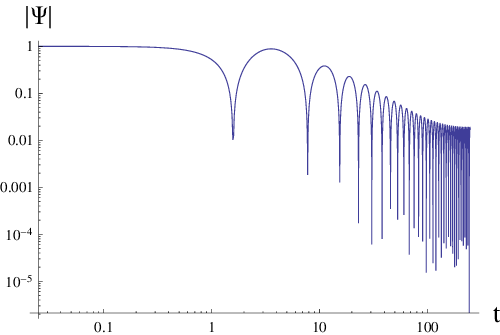}}
\caption{Black hole model I. Logarithmic time-domain profile for $\ell=1$ perturbations. The parameters are $l=0.51$, $M=1$, $\mu=1$. The intermediate oscillatory late time tails with power-law envelope dominate even in the early phase, so that the quasi-resonant modes cannot be extracted from the signal.}\label{fig:TD3}
\end{figure}

In our implementation of the time-domain integration, we follow the characteristic scheme of Gundlach, Price, and Pullin \cite{Gundlach:1993tp}, discretizing the wave equation on a uniform grid in the light-cone coordinates $(u,v)$ according to Eq.~\eqref{GPP}. The evolution is initiated by specifying the field on the initial null surfaces, where we choose a Gaussian wave packet of the form $\Psi(u=u_0,v)=\exp[-(v-v_c)^2/(2\sigma^2)]$ and $\Psi(u,v=v_0)=0$, with $v_c$ determining the initial location of the pulse and $\sigma$ its width. The overall normalization of the initial data is arbitrary due to the linearity of the perturbation equation; throughout the evolution we therefore work with dimensionless amplitudes, so that the vertical scale in Figs.~\ref{fig:TD2} and \ref{fig:TD3} reflects the relative decay of the signal rather than an absolute physical normalization. The resulting time-domain profile consists of an initial transient stage, followed by a period of exponentially damped oscillations dominated by QNMs, and finally a late-time tail, in agreement with the general expectations for black-hole perturbations. The quasinormal frequencies are extracted from the intermediate stage using the Prony method, which approximates the signal as a superposition of damped exponentials $\Psi(t)=\sum_k C_k e^{-i\omega_k t}$ and determines the complex frequencies $\omega_k$ via a linear fitting procedure (see, \cite{London:2014cma,Bolokhov:2025fto,Lutfuoglu:2025mqa,Konoplya:2022xid} for recent examples). In practice, we select a time window where the ringing stage is clearly established and the contamination from both the initial burst and the late-time tail is negligible; the stability of the extracted frequencies with respect to variations of this window provides an estimate of the numerical uncertainty.

The time-domain integration approach has been widely applied in the study of black hole perturbations. It provides a robust way to extract the dominant quasinormal mode and to reveal possible dynamical instabilities from the temporal evolution of perturbations. Owing to these advantages, the method has been implemented in numerous investigations (see, for example, \cite{Konoplya:2021ube, Konoplya:2013sba, Skvortsova:2023zca, Dubinsky:2025wns, Ishihara:2008re, Konoplya:2005et, Stuchlik:2025mjj, Dubinsky:2024gwo, Abdalla:2005hu, Cuyubamba:2016cug, Konoplya:2014lha, Lutfuoglu:2025blw, Konoplya:2023ahd, Dubinsky:2025bvf, Skvortsova:2025cah, Lutfuoglu:2025bsf, Dubinsky:2024mwd, Bolokhov:2025lnt}), demonstrating its effectiveness for determining the fundamental oscillation frequency and diagnosing instabilities.

\begin{table}
\begin{tabular}{c c c c}
\hline
\hline
$\mu$ & $WKB_{16}$ & $WKB_{14}$ & diff  $(\%)$ \\
\hline
\hline
\multicolumn{4}{c}{$\ell=0$, $4 l^4=9.4$, $n=0$} \\
\hline
$0$ & $0.105188-0.093854 i$ & $0.105228-0.093839 i$ & $0.0307$\\
$0.025$ & $0.105450-0.093341 i$ & $0.105492-0.093325 i$ & $0.0316$\\
$0.05$ & $0.106203-0.091798 i$ & $0.106255-0.091780 i$ & $0.0394$\\
$0.075$ & $0.107262-0.089156 i$ & $0.107447-0.089135 i$ & $0.133$\\
$0.1$ & $0.109222-0.086386 i$ & $0.109547-0.085847 i$ & $0.452$\\
$0.125$ & $0.111004-0.081738 i$ & $0.111116-0.081834 i$ & $0.107$\\
$0.15$ & $0.113192-0.077873 i$ & $0.114042-0.077646 i$ & $0.641$\\
$0.175$ & $0.116283-0.070686 i$ & $0.115010-0.069957 i$ & $1.08$\\
$0.2$ & $0.111256-0.066627 i$ & $0.115324-0.075083 i$ & $7.24$\\

\hline
\multicolumn{4}{c}{$\ell=1$, $4 l^4=9.4$, $n=0$} \\
\hline
$0$ & $0.291737-0.090020 i$ & $0.291738-0.090021 i$ & $0.0004$\\
$0.05$ & $0.292902-0.089469 i$ & $0.292903-0.089470 i$ & $0.0004$\\
$0.1$ & $0.296401-0.087794 i$ & $0.296401-0.087795 i$ & $0.0003$\\
$0.15$ & $0.302241-0.084928 i$ & $0.302241-0.084928 i$ & $0$\\
$0.2$ & $0.310439-0.080750 i$ & $0.310435-0.080748 i$ & $0.0013$\\
$0.25$ & $0.320980-0.075090 i$ & $0.320990-0.075092 i$ & $0.0031$\\
$0.3$ & $0.333916-0.067658 i$ & $0.333916-0.067649 i$ & $0.00253$\\
$0.35$ & $0.349107-0.058140 i$ & $0.349108-0.058129 i$ & $0.00316$\\
$0.4$ & $0.366254-0.046244 i$ & $0.366589-0.046358 i$ & $0.0958$\\
$0.45$ & $0.381614-0.033471 i$ & $0.383654-0.035630 i$ & $0.775$\\

\hline
\multicolumn{4}{c}{$\ell=2$, $4 l^4=9.4$, $n=0$} \\
\hline
$0$ & $0.484763-0.090095 i$ & $0.484763-0.090095 i$ & $0$\\
$0.05$ & $0.485547-0.089878 i$ & $0.485547-0.089878 i$ & $0$\\
$0.1$ & $0.487900-0.089223 i$ & $0.487900-0.089223 i$ & $0$\\
$0.15$ & $0.491833-0.088118 i$ & $0.491833-0.088118 i$ & $0$\\
$0.2$ & $0.497358-0.086547 i$ & $0.497358-0.086547 i$ & $0$\\
$0.25$ & $0.504497-0.084482 i$ & $0.504497-0.084482 i$ & $0$\\
$0.3$ & $0.513277-0.081888 i$ & $0.513277-0.081888 i$ & $0$\\
$0.35$ & $0.523732-0.078719 i$ & $0.523732-0.078719 i$ & $0$\\
$0.4$ & $0.535904-0.074916 i$ & $0.535904-0.074916 i$ & $0.00007$\\
$0.45$ & $0.549843-0.070403 i$ & $0.549842-0.070403 i$ & $0.00017$\\
$0.5$ & $0.565603-0.065081 i$ & $0.565603-0.065081 i$ & $0.00008$\\
$0.55$ & $0.583242-0.058818 i$ & $0.583242-0.058818 i$ & $0.00003$\\
$0.6$ & $0.602797-0.051440 i$ & $0.602808-0.051436 i$ & $0.00195$\\
$0.65$ & $0.624298-0.042671 i$ & $0.624283-0.042657 i$ & $0.00331$\\
$0.7$ & $0.647280-0.032298 i$ & $0.647238-0.032325 i$ & $0.00761$\\
\hline
\hline
\end{tabular}
\caption{Quasinormal modes of the scalar perturbations of the regular black-hole I ($M=1$) calculated using the WKB method at different orders (8th and 10th orders) with Padé approximants for $4 l^4=9.4$ and $n=0$.} \label{tableI}
\end{table}

\begin{table}
\centering
\begin{tabular}{c c c c}
\hline
\hline
$\mu$ & $WKB_{16}$ & $WKB_{14}$ & diff  $(\%)$ \\
\hline
\multicolumn{4}{c}{$\ell=0$, $n=0$, $l=0.51$}\\
\hline
$0$ & $0.220563-0.178548 i$ & $0.220603-0.178114 i$ & $0.154$\\
$0.025$ & $0.220706-0.178319 i$ & $0.220747-0.177881 i$ & $0.155$\\
$0.05$ & $0.221133-0.177631 i$ & $0.221174-0.177183 i$ & $0.159$\\
$0.075$ & $0.221827-0.176487 i$ & $0.221876-0.176021 i$ & $0.165$\\
$0.1$ & $0.222758-0.174891 i$ & $0.222837-0.174398 i$ & $0.176$\\
$0.125$ & $0.223847-0.172834 i$ & $0.224034-0.172319 i$ & $0.194$\\
$0.15$ & $0.224966-0.169963 i$ & $0.225424-0.169789 i$ & $0.174$\\
$0.175$ & $0.227237-0.166776 i$ & $0.227143-0.166777 i$ & $0.0335$\\
$0.2$ & $0.229109-0.163647 i$ & $0.228865-0.163485 i$ & $0.104$\\
$0.225$ & $0.231036-0.160045 i$ & $0.230703-0.159860 i$ & $0.135$\\
$0.25$ & $0.233352-0.156174 i$ & $0.232655-0.156286 i$ & $0.251$\\
$0.275$ & $0.235763-0.151123 i$ & $0.235817-0.152930 i$ & $0.646$\\
$0.3$ & $0.236265-0.146598 i$ & $0.240062-0.146678 i$ & $1.37$\\

\hline
\multicolumn{4}{c}{$\ell=1$, $n=0$, $l=0.51$}\\
\hline
 $0$ & $0.610557-0.167819 i$ & $0.610557-0.167819 i$ & $0$\\
 $0.1$ & $0.612667-0.166952 i$ & $0.612667-0.166952 i$ & $0$\\
 $0.2$ & $0.619007-0.164308 i$ & $0.619007-0.164308 i$ & $0$\\
 $0.3$ & $0.629609-0.159753 i$ & $0.629609-0.159753 i$ & $0$\\
 $0.4$ & $0.644520-0.153047 i$ & $0.644519-0.153046 i$ & $0.0002$\\
 $0.5$ & $0.663786-0.143824 i$ & $0.663786-0.143823 i$ & $0.0001$\\
$0.6$ & $0.687413-0.131545 i$ & $0.687413-0.131546 i$ & $0.0001$\\
 $0.7$ & $0.715262-0.115486 i$ & $0.715261-0.115485 i$ & $0.0002$\\
 $0.8$ & $0.746910-0.094855 i$ & $0.746932-0.094898 i$ & $0.0064$\\
 $0.9$ & $0.781400-0.070528 i$ & $0.781474-0.070888 i$ & $0.0468$\\

\hline
\multicolumn{4}{c}{$\ell=2$, $n=0$, $l=0.51$}\\
\hline
 $0$ & $1.012054-0.167025 i$ & $1.012054-0.167025 i$ & $0$\\
 $0.1$ & $1.013446-0.166691 i$ & $1.013446-0.166691 i$ & $0$\\
 $0.2$ & $1.017630-0.165682 i$ & $1.017630-0.165682 i$ & $0$\\
 $0.3$ & $1.024626-0.163977 i$ & $1.024626-0.163977 i$ & $0$\\
 $0.4$ & $1.034466-0.161539 i$ & $1.034466-0.161539 i$ & $0$\\
 $0.5$ & $1.047200-0.158316 i$ & $1.047200-0.158316 i$ & $0$\\
$0.6$ & $1.062891-0.154236 i$ & $1.062891-0.154236 i$ & $0$\\
 $0.7$ & $1.081619-0.149204 i$ & $1.081619-0.149204 i$ & $0$\\
 $0.8$ & $1.103482-0.143097 i$ & $1.103482-0.143097 i$ & $0$\\
 $0.9$ & $1.128596-0.135756 i$ & $1.128596-0.135756 i$ & $0$\\
 $1.0$ & $1.157094-0.126970 i$ & $1.157094-0.126970 i$ & $0$\\
 $1.1$ & $1.189117-0.116458 i$ & $1.189117-0.116458 i$ & $0$\\
 $1.2$ & $1.224790-0.103842 i$ & $1.224789-0.103842 i$ & $0$\\
 $1.3$ & $1.264169-0.088633 i$ & $1.264173-0.088648 i$ & $0.0012$\\
 $1.4$ & $1.307125-0.070261 i$ & $1.307130-0.070363 i$ & $0.0078$\\
$1.5$ & $1.385769-0.012815 i$ & $1.387581-0.010364 i$ & $0.220$\\
\hline
\hline
\end{tabular}
\caption{Fundamental ($n=0$) QNMs of the scalar potential for the regular black-hole II with $M=1$ and $l=0.51$. Results are obtained using WKB formulas with Padé approximants at different orders. The last column shows the relative deviation between the two WKB orders.} \label{tableII}
\end{table}

\begin{table*}
\begin{tabular}{l c c c c c c c}
\hline
\hline
$4 l^4$ & $r_0$ & $T_H$ & $r_{m}$ & $R_{s}$ & $\lambda$ & $\Omega_{ISCO}$ & BE\\
\hline
$1.$ & $1.984$ & $0.03817$ & $2.994$ & $5.19256$ & $0.191377$ & $0.06807$ & $0.05720$\\
$1.4$ & $1.977$ & $0.03749$ & $2.991$ & $5.19111$ & $0.190939$ & $0.06809$ & $0.05721$\\
$1.8$ & $1.970$ & $0.03678$ & $2.989$ & $5.18965$ & $0.190494$ & $0.06810$ & $0.05721$\\
$2.2$ & $1.963$ & $0.03604$ & $2.986$ & $5.18818$ & $0.190044$ & $0.06812$ & $0.05721$\\
$2.6$ & $1.955$ & $0.03527$ & $2.983$ & $5.18671$ & $0.189587$ & $0.06813$ & $0.05722$\\
$3.$ & $1.947$ & $0.03447$ & $2.981$ & $5.18522$ & $0.189124$ & $0.06814$ & $0.05722$\\
$3.4$ & $1.939$ & $0.03364$ & $2.978$ & $5.18372$ & $0.188655$ & $0.06816$ & $0.05723$\\
$3.8$ & $1.930$ & $0.03276$ & $2.976$ & $5.18222$ & $0.188180$ & $0.06817$ & $0.05723$\\
$4.2$ & $1.921$ & $0.03184$ & $2.973$ & $5.18070$ & $0.187697$ & $0.06818$ & $0.05723$\\
$4.6$ & $1.912$ & $0.03088$ & $2.970$ & $5.17918$ & $0.187208$ & $0.06820$ & $0.05724$\\
$5.$ & $1.902$ & $0.02985$ & $2.967$ & $5.17765$ & $0.186712$ & $0.06821$ & $0.05724$\\
$5.4$ & $1.892$ & $0.02876$ & $2.965$ & $5.17610$ & $0.186208$ & $0.06822$ & $0.05725$\\
$5.8$ & $1.880$ & $0.02760$ & $2.962$ & $5.17455$ & $0.185697$ & $0.06824$ & $0.05725$\\
$6.2$ & $1.869$ & $0.02634$ & $2.959$ & $5.17298$ & $0.185178$ & $0.06825$ & $0.05725$\\
$6.6$ & $1.856$ & $0.02498$ & $2.956$ & $5.17141$ & $0.184651$ & $0.06827$ & $0.05726$\\
$7.$ & $1.842$ & $0.02349$ & $2.953$ & $5.16982$ & $0.184116$ & $0.06828$ & $0.05726$\\
$7.4$ & $1.826$ & $0.02183$ & $2.950$ & $5.16822$ & $0.183572$ & $0.06829$ & $0.05727$\\
$7.8$ & $1.809$ & $0.01994$ & $2.947$ & $5.16661$ & $0.183020$ & $0.06831$ & $0.05727$\\
$8.2$ & $1.788$ & $0.01775$ & $2.944$ & $5.16499$ & $0.182458$ & $0.06832$ & $0.05727$\\
$8.6$ & $1.764$ & $0.01507$ & $2.941$ & $5.16336$ & $0.181888$ & $0.06834$ & $0.05728$\\
$9.$ & $1.732$ & $0.01149$ & $2.938$ & $5.16171$ & $0.181307$ & $0.06835$ & $0.05728$\\
$9.4$ & $1.675$ & $0.00500$ & $2.935$ & $5.16005$ & $0.180717$ & $0.06836$ & $0.05729$\\
\hline
\hline
\end{tabular}
\caption{Event horizon position ($r_0$), Hawking temperature ($T_H$), photon-sphere radius ($r_m$), shadow radius ($R_s$), Lyapunov exponent ($\lambda$), ISCO frequency ($\Omega_{ISCO}$) and binding energy (BE) for the regular black hole I ($M=1$).} \label{tableIII}
\end{table*}

\begin{table*}
\begin{tabular}{l c c c c c c c}
\hline
\hline
$l$ & $r_0$ & $T_H$ & $r_{m}$ & $R_{s}$ & $\lambda$ & $\Omega_{ISCO}$ & BE\\
\hline
$0.04$ & $0.999$ & $0.07945$ & $1.499$ & $2.59746$ & $0.384717$ & $0.13614$ & $0.05720$\\
$0.08$ & $0.997$ & $0.07906$ & $1.497$ & $2.59560$ & $0.384163$ & $0.13632$ & $0.05725$\\
$0.12$ & $0.993$ & $0.07840$ & $1.494$ & $2.59249$ & $0.383221$ & $0.13661$ & $0.05732$\\
$0.16$ & $0.987$ & $0.07743$ & $1.488$ & $2.58808$ & $0.381861$ & $0.13703$ & $0.05742$\\
$0.2$ & $0.979$ & $0.07612$ & $1.482$ & $2.58232$ & $0.380039$ & $0.13758$ & $0.05755$\\
$0.24$ & $0.969$ & $0.07441$ & $1.473$ & $2.57513$ & $0.377690$ & $0.13827$ & $0.05771$\\
$0.28$ & $0.957$ & $0.07218$ & $1.463$ & $2.56643$ & $0.374724$ & $0.13910$ & $0.05790$\\
$0.32$ & $0.941$ & $0.06930$ & $1.451$ & $2.55609$ & $0.371010$ & $0.14009$ & $0.05813$\\
$0.36$ & $0.922$ & $0.06549$ & $1.436$ & $2.54392$ & $0.366364$ & $0.14125$ & $0.05840$\\
$0.4$ & $0.897$ & $0.06030$ & $1.418$ & $2.52971$ & $0.360509$ & $0.14261$ & $0.05871$\\
$0.44$ & $0.864$ & $0.05269$ & $1.397$ & $2.51313$ & $0.353016$ & $0.14419$ & $0.05907$\\
$0.48$ & $0.814$ & $0.03950$ & $1.372$ & $2.49371$ & $0.343173$ & $0.14603$ & $0.05948$\\
\hline
\hline
\end{tabular}
\caption{Event horizon position ($r_0$), Hawking temperature ($T_H$), photon-sphere radius ($r_m$), shadow radius ($R_s$), Lyapunov exponent ($\lambda$), ISCO frequency ($\Omega_{ISCO}$) and binding energy (BE) for the regular black hole ($M=1$).} \label{tableIV}
\end{table*}

\section{Quasinormal modes}\label{sec:SecV}

The numerical results for the fundamental QNMs of a massive scalar field are presented in Tables~\ref{tableI} and ~\ref{tableII} for the two regular black–hole geometries considered in this work. Several characteristic features of the spectrum can be identified from these data.

First, the real part of the frequency generally increases with the mass of the scalar field. This behavior is clearly visible for all multipole numbers $\ell$ shown in Tables~\ref{tableI} and ~\ref{tableII}. At the same time, the imaginary part of the frequency steadily decreases in magnitude as $\mu$ grows. Since the damping rate is determined by $|\operatorname{Im}\omega|$, this implies that the oscillations become progressively longer lived when the scalar-field mass increases. For sufficiently large $\mu$, the imaginary part approaches zero, which is most evident for higher multipoles. For example, for $\ell=2$ in Table~\ref{tableII} the imaginary part decreases from $0.167025$ at $\mu=0$ to $0.012815$ at $\mu=1.5$. This tendency indicates the approach to the regime of quasi-resonances, in which the damping time becomes arbitrarily large. Although the present calculations do not reach the exact quasi-resonant point, extrapolation of the trend observed in the tables strongly suggests that such modes should occur at slightly larger values of $\mu$.

Second, the comparison of different WKB orders demonstrates that the method provides highly stable results when sufficiently high orders and Padé resummation are used. The relative differences between the 16th- and 14th-order WKB values are extremely small in most of the parameter range. For $\ell=1$ and $\ell=2$ the deviation typically lies in the range $10^{-4}$--$10^{-2}\%$, indicating excellent convergence of the WKB expansion. Even in the least favorable cases, the discrepancy remains well below one percent. Such an agreement between adjacent WKB orders serves as a practical indicator of the reliability of the approximation.

Finally, the magnitude of the physical effect caused by the scalar-field mass significantly exceeds the expected numerical uncertainty of the WKB method. The variation of the quasinormal frequencies with $\mu$ changes the oscillation frequency and damping rate by many percent across the considered parameter range, while the relative numerical error inferred from the difference between neighboring WKB orders is typically orders of magnitude smaller. Therefore, the observed dependence of the spectrum on the field mass represents a genuine physical effect rather than a numerical artifact.

Overall, the results show that the regular black-hole geometries considered here support the same qualitative behavior of massive scalar perturbations that is known for many other black-hole backgrounds: increasing the field mass leads to weaker damping and the eventual appearance of long-lived modes. The data presented in Tables~\ref{tableI} and ~\ref{tableII} provide clear indications that the spectrum approaches the quasi-resonant regime as $\mu$ increases.

An additional consistency check was performed using time-domain integration. For moderate values of the scalar-field mass the time-domain profiles show a clear stage of exponentially damped oscillations dominated by the fundamental QNM. In this regime, the frequencies extracted from the profiles are in excellent agreement with the WKB results obtained with Padé resummation, as illustrated in Fig.~\ref{fig:TD2}, where the dominant frequencies determined from the time-domain signal practically coincide with the WKB values. For larger values of the field mass, however, the structure of the time-domain signal changes qualitatively. The onset of the asymptotic stage occurs earlier, and the late-time behavior becomes dominated by oscillatory tails with a power-law envelope (see Fig.~\ref{fig:TD3}). As a result, the quasi-resonant mode expected from the extrapolation of the WKB spectrum does not manifest itself as a distinct stage in the time-domain profile, being effectively masked by the early appearance of the oscillatory tails. This behavior is consistent with the known properties of massive-field perturbations, for which oscillatory power-law tails dominate the asymptotic signal.

Notice that the obtained QNMs can be used to obtain the grey-body factors $\Gamma_{\ell}(\Omega)$ of a massive scalar field via the correspondence developed in \cite{Konoplya:2024vuj} 
\small
\begin{eqnarray}\nonumber
\Gamma_{\ell}(\Omega)\equiv |T|^2 &=&
\left(1+e^{2\pi\dfrac{\Omega^2-\re{\omega_0}^2}{4\re{\omega_0}\im{\omega_0}}}\right)^{-1}+ \Order{\ell^{-1}} .
\end{eqnarray} \normalsize
and tested in numerous publications \cite{Skvortsova:2024msa, Bolokhov:2024otn, Malik:2024cgb, Malik:2025dxn, Dubinsky:2025nxv}.

Although gravitational perturbations generally differ from those of test fields, numerous studies have shown that their spectra share qualitatively similar features, especially at moderate and high multipole numbers. This is because the dominant contribution to the effective potential is given by the centrifugal term $\ell(\ell+1)/r^{2}$, which quickly outweighs subleading corrections, so that the test-field approximation frequently captures the main characteristics of the spectrum.

\section{Particle motion}\label{sec:SecVI}

To analyze geodesic motion in the spacetime under consideration, it is convenient to introduce the auxiliary function \cite{Konoplya:2020hyk}
\begin{equation}\label{Pdef}
P(r)=\frac{f(r)}{r^{2}}.
\end{equation}
This function plays an important role in identifying the region where classical radiation processes are most relevant. Roughly speaking, the so-called radiation zone corresponds to the region surrounding the maximum of $P(r)$, while the near-horizon region and the asymptotic domain far from the ISCO do not contribute significantly to classical radiation effects.

The four-momentum of a particle of mass $m$ moving along a geodesic is defined as
\begin{equation}\label{pmu}
p^{\mu}=m\frac{dx^{\mu}}{ds},
\end{equation}
where $s$ is the affine parameter along the trajectory. Owing to the stationarity and spherical symmetry of the spacetime, the energy and angular momentum are conserved quantities,
\begin{equation}
E=-p_{t}, \qquad L=p_{\phi}.
\end{equation}
The normalization condition for the four-momentum,
\begin{equation}\label{norm}
p_{\mu}p^{\mu}=-m^{2},
\end{equation}
leads to the radial equation of motion.

Without loss of generality, the motion can be restricted to the equatorial plane $\theta=\pi/2$, so that $d\theta=0$. In this case, the radial dynamics can be written in the form
\begin{equation}\label{radialeq2}
m^{2}\frac{1}{f(r)}\left(\frac{dr}{ds}\right)^{2}=V_{\mathrm{eff}}(r),
\end{equation}
where the effective potential is
\begin{equation}\label{Veff}
V_{\mathrm{eff}}(r)=\frac{E^{2}}{f(r)}-\frac{L^{2}}{r^{2}}-m^{2}.
\end{equation}

Circular geodesics correspond to a constant radial coordinate. This condition is equivalent to
\begin{equation}\label{circcond}
V_{\mathrm{eff}}(r)=0, \qquad V_{\mathrm{eff}}'(r)=0.
\end{equation}
Solving these relations for the conserved quantities yields
\begin{equation}\label{ELcirc}
\begin{aligned}
E^{2} &= -m^{2}\frac{2rP^{2}(r)}{P'(r)}, \\
L^{2} &= -m^{2}\frac{f'(r)}{P'(r)}.
\end{aligned}
\end{equation}

The angular velocity of a particle moving along a circular orbit is therefore
\begin{equation}\label{Omega}
\Omega^{2}\equiv\left(\frac{d\phi}{dt}\right)^{2} =\frac{L^{2}f^{2}(r)}{E^{2}r^{4}}=\frac{f'(r)}{2r}.
\end{equation}

\subsection{Photon sphere and shadow}

Black-hole shadows provide a direct observational probe of the geometry in the strong-field regime. The first horizon-scale images of supermassive black holes obtained by the Event Horizon Telescope have opened a new possibility to test gravitational theories and constrain deviations from the Schwarzschild and Kerr geometries \cite{EventHorizonTelescope:2019dse, EventHorizonTelescope:2022wkp}. Because the shadow size and shape are determined by the properties of photon orbits near the photon sphere, they are particularly sensitive to modifications of the spacetime metric in the vicinity of the black hole \cite{Perlick:2015vta}.

For massless particles ($m=0$), the circular orbit is determined by the condition that $P(r)$ reaches its maximum. Denoting the corresponding radius by $r_{m}$, one finds
\begin{equation}
P'(r_m)=0 .
\end{equation}
The radius of the black-hole shadow is determined by the critical impact parameter $R_s=L/E$. Using Eq.~(\ref{ELcirc}) one obtains
\begin{equation}\label{shadow2}
\frac{1}{R_s^2}=P(r_m).
\end{equation}

The instability of photon orbits can be characterized by the Lyapunov exponent. Introducing a small radial perturbation $r=r_m+\delta r$ and expanding the equation of motion to leading order gives
\begin{equation}\label{lyeq}
\left(\frac{d}{dt}\delta r\right)^{2}=\lambda^{2}\delta r^{2}+\mathcal{O}(\delta r^{3}),
\end{equation}
where the Lyapunov exponent is
\begin{equation}\label{lyapunov}
\lambda^{2}=-\frac{1}{2 P}\frac{d^{2}P}{dr_*^{2}}\Bigg|_{r=r_m}.
\end{equation}
This quantity measures the rate at which nearby photon trajectories diverge and therefore determines the instability timescale of the photon sphere.

\subsection{Innermost stable circular orbit}

For massive particles, stability of circular motion is determined by the second derivative of the effective potential. Expanding the radial equation near the circular orbit yields
\begin{equation}
m^{2}\frac{1}{f(r)}\left(\frac{d}{ds}\delta r\right)^{2}=\frac{V_{\mathrm{eff}}''(r)}{2}\delta r^{2}+\mathcal{O}(\delta r^{3}).
\end{equation}
Stable circular orbits correspond to
\begin{equation}
V_{\mathrm{eff}}''(r)<0 .
\end{equation}

In asymptotically flat spacetimes, the marginal stability condition
\begin{equation}
V_{\mathrm{eff}}''(r)=0
\end{equation}
defines the radius of the innermost stable circular orbit $r_{\mathrm{ISCO}}$. The corresponding energy is obtained from Eq.~(\ref{ELcirc}) as
\begin{equation}
E_{\mathrm{ISCO}}^{2}=-m^{2}\max\frac{2rP^{2}(r)}{P'(r)} .
\end{equation}

Two important invariant characteristics of the ISCO are its angular frequency
\begin{equation}
\Omega_{\mathrm{ISCO}}=\Omega(r_{\mathrm{ISCO}})
\end{equation}
and the binding energy of a particle moving from a distant orbit to the ISCO,
\begin{equation}\label{binding}
BE=1-\frac{E_{\mathrm{ISCO}}}{m}.
\end{equation}
The binding energy represents the fraction of the particle's rest mass that can be released through accretion processes in the vicinity of the black hole.

\subsection{Particle motion characteristics}

The numerical results for the properties of circular geodesics and photon motion are summarized in Tables~\ref{tableIII} and ~\ref{tableIV} for the two regular black-hole geometries. These quantities provide a complementary probe of the spacetime structure and allow one to assess how the regularization parameter modifies the strong-field region of the geometry.

For the first model, the variation of the parameter controlling the deviation from the Schwarzschild solution produces only moderate changes in the characteristic radii and dynamical quantities associated with photon motion. As shown in Table~\ref{tableIII}, the radius of the photon sphere $r_m$ decreases slightly as the parameter $4l^{4}$ increases, while the corresponding shadow radius $R_s$ also exhibits a small monotonic decrease. Over the entire parameter range considered, the change of these quantities is relatively small, indicating that the geometry remains close to the Schwarzschild case in the region relevant for photon trajectories. The Lyapunov exponent governing the instability of photon orbits decreases gradually as well, implying that photon trajectories become marginally less unstable when the deviation from the Schwarzschild geometry
increases.

The quantities associated with massive particle motion exhibit a similarly mild dependence on the parameter of the solution. The angular frequency at the ISCO varies only slightly across the parameter range, while the binding energy remains nearly constant. This behavior suggests that the efficiency of accretion processes around these regular black holes is only weakly affected by the modification of the metric function.

For the second model, the influence of the parameter $l$ on the geodesic structure is more pronounced. As seen in Table~\ref{tableIV}, increasing $l$ leads to a noticeable decrease in the event-horizon radius and the Hawking temperature. At the same time, the radius of the photon sphere and the shadow radius both decrease monotonically as the parameter grows. The Lyapunov exponent also decreases, indicating a gradual reduction in the instability of photon orbits.

An interesting feature revealed by the data in Table~\ref{tableIV} is the appearance of stable circular orbits only beyond a certain value of the parameter. For sufficiently small $l$, no ISCO is found in the range explored, whereas for larger values of $l$, the ISCO frequency and the corresponding binding energy can be defined. Once the ISCO exists, both quantities increase slowly with increasing $l$, indicating that circular particle motion becomes slightly more tightly bound in the stronger regularization regime.

Overall, the results show that the modifications introduced by the quasi-topological gravity corrections affect the geodesic structure predominantly through small shifts of the characteristic radii and frequencies. The photon sphere, the shadow radius, and the Lyapunov exponent change only moderately, while the parameters associated with massive particle motion remain close to their Schwarzschild values within the considered range of the regularization parameter.

\section{Conclusions} \label{sec:SecVII}

While various properties, perturbations, and spectra of black holes in theories with higher-curvature corrections have been extensively studied (see, for instance, \cite{Moura:2006pz, Prasobh:2014zea, Grozdanov:2016fkt, Konoplya:2020ibi, Gonzalez:2018xrq, Konoplya:2017zwo, Cano:2020cao, Blazquez-Salcedo:2020caw, Blazquez-Salcedo:2020rhf} and references therein), only a few recent works have been devoted to regular black holes arising in such theories. Here, we have studied QNMs of a massive scalar field and several characteristics of particle motion in the spacetime of regular black holes
arising in four-dimensional quasi-topological gravity.

The QNM spectrum exhibits a clear dependence on the scalar-field mass. As $\mu$ increases, the oscillation frequency grows while the damping rate decreases substantially. This tendency indicates the approach to the regime of quasi-resonances, in which the lifetime of perturbations becomes very long. The comparison of high WKB orders with Padé resummation shows excellent convergence, and the differences between adjacent orders are much smaller than the changes in the spectrum produced by varying the field mass. This confirms the reliability of the obtained frequencies within the explored parameter range.

We have also analyzed several invariant characteristics of particle motion. For the first regular black-hole model, the photon-sphere radius, the shadow size, the Lyapunov exponent, the ISCO frequency, and the binding energy change only slightly with the regularization parameter, indicating that the spacetime remains close to the Schwarzschild geometry in the region relevant for geodesic motion. In the second model, the influence of the parameter $l$ is somewhat stronger, leading to monotonic shifts of the photon-sphere and shadow radii and a gradual decrease of the Lyapunov exponent. Stable circular orbits appear only beyond a certain value of the parameter, after which the ISCO frequency and the binding energy increase slowly.

Overall, we observe that observables related to particle motion and the least damped QNMs exhibit only a mild dependence on the coupling parameter responsible for regularization. In contrast, quantities determined by the near-horizon geometry, such as the Hawking temperature, can vary significantly with the coupling.  This behavior is a common feature of various regular black hole models, including those of Hayward, Dymnikova, and Bardeen. It is apparently related to the fact that the least damped modes, as well as key characteristics of particle motion, are predominantly governed by the geometry in the region between the photon sphere and the innermost stable circular orbit, i.e., at a finite distance from the event horizon.  At the same time, these regular black hole solutions exhibit substantial deviations from the Schwarzschild geometry in the immediate vicinity of the horizon, while rapidly approaching the Schwarzschild limit at larger radial distances. The results demonstrate that massive scalar perturbations in these spacetimes naturally approach the long-lived quasi-resonant regime, providing a characteristic signature of massive-field dynamics around regular black holes.

\begin{acknowledgments}
B. C. L. is grateful to the Excellence project FoS UHK 2205/2025-2026 for the financial support.
\end{acknowledgments}

\bibliography{bibliography}
\end{document}